%% file: main.tex
\documentclass[sigconf]{acmart}

\usepackage{amsmath}
\usepackage{soul}
\usepackage{todonotes}

\usepackage{xcolor}
\definecolor{entropy_blue}{RGB}{35, 119, 181}
\definecolor{info_gain_orange}{RGB}{242, 105, 18}
\definecolor{low_ig}{RGB}{233,174,143}
\definecolor{low_entropy}{RGB}{124,176,222}

\definecolor{pumpkin}{RGB}{211, 84, 0}

\usepackage{multirow}
\usepackage{float}
\usepackage{graphicx}
\usepackage{longtable}
\usepackage[justification=centering]{subfig}
\usepackage{makecell}
\usepackage{booktabs}
\usepackage{arydshln}
\usepackage{tabularx}
\usepackage{stfloats}

\copyrightyear{2024}
\acmYear{2024}
\setcopyright{rightsretained}
\acmConference[C\&C '24]{Creativity and Cognition}{June 23--26, 2024}{Chicago, IL, USA}
\acmBooktitle{Creativity and Cognition (C\&C '24), June 23--26, 2024, Chicago, IL, USA}
\acmDOI{10.1145/3635636.3664260}
\acmISBN{979-8-4007-0485-7/24/06}

\begin{document}

\title{CoCo Matrix: Taxonomy of Cognitive Contributions in Co-writing with Intelligent Agents}

\author{Ruyuan Wan}
\affiliation{%
  \institution{University of Notre Dame}
  \city{Notre Dame}
  \country{USA}}
\email{rwan@nd.edu}

\author{Simret Gebreegziabher}
\affiliation{%
  \institution{University of Notre Dame}
  \city{Notre Dame}
  \country{USA}}
\email{sgebreeg@nd.edu}

\author{Toby Jia-Jun Li}
\affiliation{%
  \institution{University of Notre Dame}
  \city{Notre Dame}
  \country{USA}}
\email{toby.j.li@nd.edu}

\author{Karla Badillo-Urquiola }
\affiliation{%
  \institution{University of Notre Dame}
  \city{Notre Dame}
  \country{USA}}
\email{kbadill3@nd.edu}

\begin{abstract}
In recent years, there has been a growing interest in employing intelligent agents in writing. Previous work emphasizes the evaluation of the quality of end product---whether it was coherent and polished, overlooking the journey that led to the product, which is an invaluable dimension of the creative process. To understand how to recognize human efforts in co-writing with intelligent writing systems, we adapt Flower and Hayes' cognitive process theory of writing and propose CoCo Matrix, a two-dimensional taxonomy of entropy and information gain, to depict the new human-agent co-writing model. We define four quadrants and situate thirty-four published systems within the taxonomy. Our research found that low entropy and high information gain systems are under-explored, yet offer promising future directions in writing tasks that benefit from the agent's divergent planning and the human's focused translation. CoCo Matrix, not only categorizes different writing systems but also deepens our understanding of the cognitive processes in human-agent co-writing. By analyzing minimal changes in the writing process, CoCo Matrix serves as a proxy for the writer's mental model, allowing writers to reflect on their contributions. This reflection is facilitated through the measured metrics of information gain and entropy, which provide insights irrespective of the writing system used.


\end{abstract}

\ccsdesc[300]{Human-centered computing~HCI theory, concepts and models}

\keywords{Interaction Paradigms, Collaborative Interaction, Creativity, Writing}

\maketitle

\input{01-introduction}

\input{02-related}

\input{03-method}

\input{05-taxonomy}

\input{06-Discussion}

\begin{acks}

This work was supported in part by a Notre Dame-IBM Technology Ethics Lab Award, an AnalytiXIN Faculty Fellowship, and a Google Research Scholar Award. Any opinions, findings, and conclusions or recommendations expressed in this material are those of the authors and do not necessarily reflect the views of the sponsors. 
\end{acks}

\bibliographystyle{ACM-Reference-Format}
\bibliography{refrence}


\end{document}

%% file: 01-introduction.tex
\section{Introduction}



The use of Artificial Intelligence~(AI) in writing, particularly through large language models~(LLMs), has been an emerging trend in computational creativity~\cite{alsharhannatural}. The act of writing, traditionally seen as an individual's unique expression, is not just the end product, but a journey marked by complex cognitive processes, choices, and iterations \cite{flower1981cognitive}. Thus, understanding both the journey and destination of writing is crucial in areas like designing writing assistant tools and writing pedagogy.


Creativity theories~\cite{hocevar1980intelligence, brown1973empirical, woodman1990interactionist}, highlight the complexities of creative processes from problem-solving to immersive activities. In education, assessing creativity focuses on competencies and innovative thinking, prioritizing thought processes over products~\cite{aacu_value}. This complexity extends to human-agent writing systems. While the evaluation of most of these writing systems relies on user's self-reported ratings, we argue that these methods fail to capture the depth of engagement and learning outcomes \cite{deslauriers2019measuring}.

Intellectual contributions in writing can be segmented into planning, translating, and reviewing phases, following Flower's cognitive process theory of writing \cite{flower1981cognitive}. Our study used a structured review in three steps to assess contributions in human-agent co-writing. We examined Natural Language Generation (NLG) and writing assistant systems to identify evaluation gaps. While some systems depend on user self-assessment and others on analyzing the final text, both approaches overlook the contributions and collaborations occurring during the writing process. To address this gap, we propose \textbf{CoCo matrix}, a taxonomy of cognitive contributions in co-writing with
intelligent agents, inspired by concepts from information science and the cognitive process theory of writing.   
The taxonomy identified \textcolor{entropy_blue}{\textbf{entropy}} and \textcolor{info_gain_orange}{\textbf{information gain}} as primary metrics. These choices were motivated by their potential to offer more nuanced insights into the dynamics between human writers and intelligent systems. Finally, we applied \textbf{CoCo matrix} to classify existing writing assistant technologies. Through this step, we assessed the systems' current state, their fit with writer needs, and identified future research directions in human-agent writing systems.

In summary, \textbf{CoCo matrix} conceptualizes the contributions of human users and AI agents as follows:

\begin{itemize}
    \item \textbf{\textcolor{entropy_blue}{High Entropy}, \textcolor{info_gain_orange}{High Information Gain}}: This category usually sees writers taking the lead in the planning phase, with the agent then translating these plans into text.
    
    \item \textbf{\textcolor{entropy_blue}{High Entropy}, \textcolor{low_ig}{Low Information Gain}}: Systems under this classification start with a base provided by the writer, with the agent then contributing by expanding the range of ideas, enhancing creative breadth.

    \item \textbf{\textcolor{low_entropy}{Low Entropy}, \textcolor{info_gain_orange}{High Information Gain}}: Here, the agent dominates the planning stage, supporting creative directions, while the human writer focuses on crafting preferred ideas into text.

    \item \textbf{\textcolor{low_entropy}{Low Entropy}, \textcolor{low_ig}{Low Information Gain}}: In this scenario, agent's role is to offer specific feedback, with the writer primarily involved in refining and revising, leading to a detail-focused collaboration.

\end{itemize}

%% file: 02-related.tex
\section{Background}

\subsection{Cognitive Process Theory of Writing}
The cognitive process theory of writing~\cite{flower1981cognitive} offers an instrumental framework to understand the nuances of human writing, which has often been overlooked in the NLG domain. This theory describes the cognitive stages and knowledge evolution from human intellect to the produced text, a process now shared with intelligent machines. It dissects writing into three main components: the task environment (the rhetorical problem and the text produced), the writer’s long-term memory (knowledge of the subject, audience, and writing plan), and the writing process itself, which includes planning, translating ideas into text, and reviewing. Planning involves setting goals and organizing ideas, translating is about turning these ideas into coherent narratives, and reviewing is an adaptive phase where revisions and evaluations refine the writing. 
As writing evolves with AI's integration, assessing the output often overshadows the collaborative journey, focusing on the product rather than the process. Yet, understanding the nuanced interplay between human and AI during the writing process is crucial for acknowledging human contributions in co-created texts. This shift in perspective from the final written product to the writing process emphasizes the importance of the creative journey, advocating for a more nuanced evaluation of human-AI writing collaborations~\cite{gero-etal-2022-design,bhat2023interacting_HIHE}. 

CoCo matrix contributes to filling the gap in a more process-focused approach to characterizing and evaluating human-AI writing collaborations by incorporating the cognitive process theory of writing. Specifically, there is a nuanced difference in the mental model of the writer and how the written text changes through the different stages of the writing process. From the writer's perspective the writing action can potentially changes the rhetorical problem, the writer’s knowledge of the subject, audience, and the overall writing plan. CoCo matrix measures the entropy and information gain of intermediate written content as a proxy to understand the change in the writer's contributions, providing insight into the mental models of the writers through the artifact.


\subsection{Intelligent Writing Assistant Tools}
Rapid advancements in NLG have significantly expanded the capabilities of intelligent writing assistant tools, accommodating a wide range of writing tasks, from creative writing~\cite{yuan2022wordcraft, chung2022talebrush, suh2024structured, yang2022ai} to argumentative writing~\cite{bao-etal-2022-aeg, liu-etal-2023-predicting}. The design of user interfaces that resonate with the mental models of users has also been a key area of focus, facilitating a more intuitive writing process \cite{dang2023choice, buschek2021impact,wan2022user}. \citet{lee2024design} categorized user interactions with these tools into three dimensions: explicit control, where users direct system behavior through interface elements~\cite{clark2021choose,chung2022talebrush}; implicit control, where the system responds to user’s text inputs~\cite{lee2022coauthor, dang2022beyond}; and no control, with predefined system behavior~\cite{jakesch2023co}. This classification highlights the trade-off between user agency and the seamless integration of intelligent agents in writing. 

The debates between Ben Shneiderman and Pattie Maes at IUI ’97 and CHI ’97 on direct manipulation versus interface agents mirror this trade-off, discussing whether users should maintain complete control over their interactions with digital interfaces or delegate certain decisions to intelligent agents~\cite{shneiderman1997direct}. Shneiderman supported direct manipulation for its benefits in learning and efficiency, while Maes pointed out its limitations, especially for users without technical skills, suggesting tasks could be partly handed over to agents. This ongoing dialogue underscores a fundamental HCI concern: finding the optimal balance between writer autonomy and agent support to enhance the writing experience.

Balancing user agency in evolving tasks is particularly difficult. As the mental models of users evolve, their desire for control over interactions with intelligent systems also changes~\cite{andrews2023role}. CoCo matrix contributes to the design of writing assistants by measuring the change in the user's interactions through the writing process.


\subsection{Evaluation of Creative Natural Language Generation Systems }
Evaluating NLG systems presents challenges given the variability of inputs and the breadth of possible outputs~\cite{gatt2018survey}. In this context, evaluations often employ both intrinsic and extrinsic methods to capture the system performance in a comprehensive way~\cite{jones1995evaluating}. Intrinsic evaluations focus on the quality of the generated text such as fluency, relevance, and correctness, often utilizing human assessments or corpus-based metrics such as BLEU~\cite{papineni2002bleu} and ROUGE~\cite{lin2004rouge}. Extrinsic evaluations, in contrast, assess the effectiveness of the text in achieving specific goals, examining the impact on end-users and the system's function in practical applications.

Despite these methods, the evaluation of NLG systems frequently centers on the final product, using rankings or user preferences~\cite{hamalainen-alnajjar-2021-human}. Recent trends emphasize evaluating the writing process to include writer behavior and cognitive contributions~\cite{gero-etal-2022-design}, highlighting the importance of understanding AI co-writing's interactive and cognitive aspects, crucial for assessing these systems in creative contexts. To that end, CoCo matrix provides a theoretical foundation and informs the design of new evaluation method, which can be intrinsic given the information gain and entropy measurement on the changed text and extrinsic in terms of the change of writer's mental model. 

%% file: 03-method.tex
\section{Methodology}

\begin{figure*}
    \centering
    \includegraphics[width=0.9\textwidth]{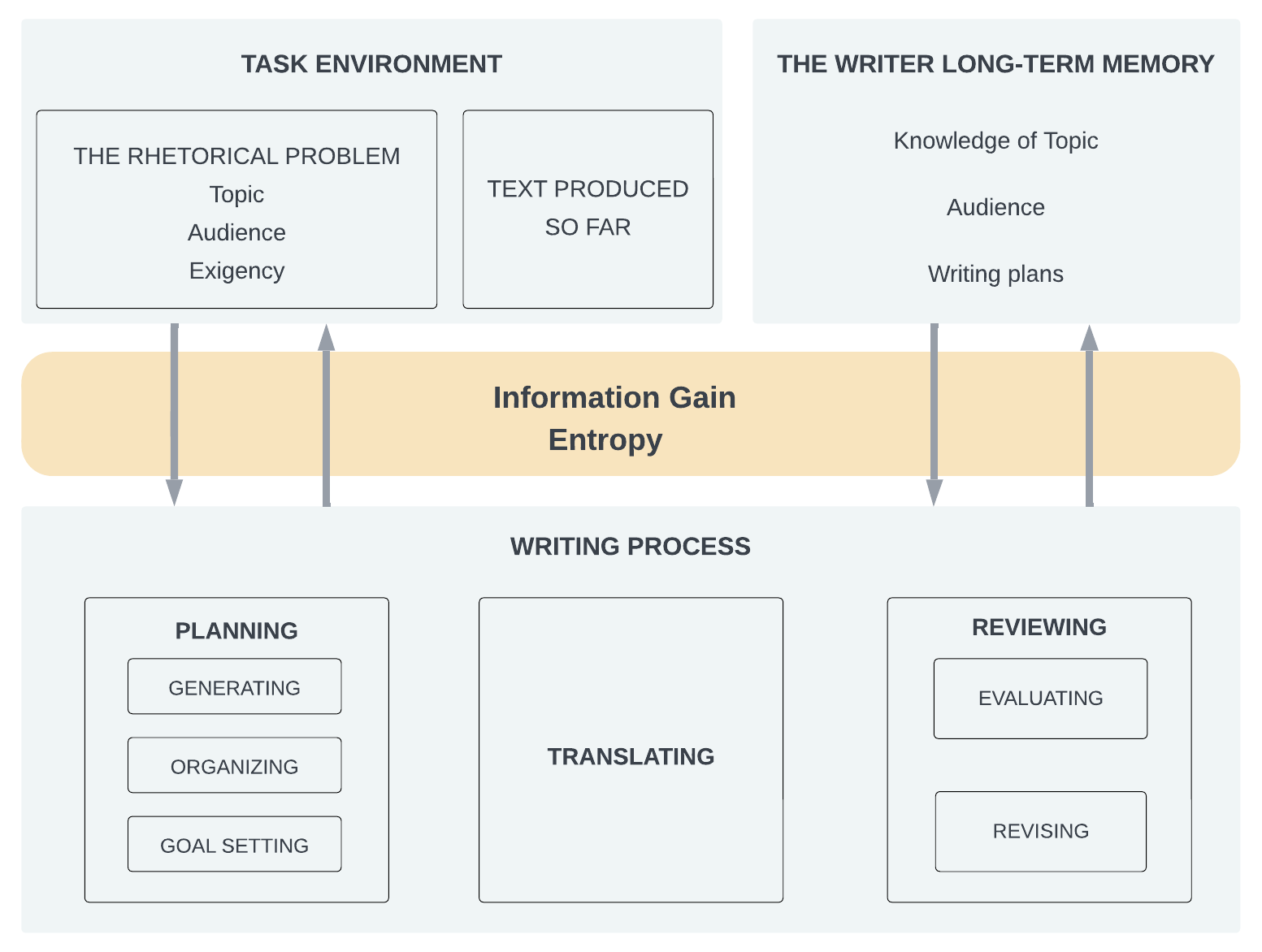}
    \caption{Adapted Cognitive Process Theory of Writing: The cyan color frames are the original structure of the writing model from Flower et al.\cite{flower1981cognitive}, we propose use entropy and information gain to depict the new model of human-agent co-writing. }
    \label{fig:updated cognitive process}
\end{figure*}


\subsection{Literature Review}
\label{sec:lit_review}
To understand human interaction with intelligent writing assistants, we conducted a literature review focusing on HCI and NLP publications from the past decade. Although engaging with a broader range of literature, including non-computing sources, would enrich our analysis, the scope of this project was primarily focused on the interaction mechanism in language technologies. Our search employed keywords such as ``creative'', ``creativity'', ``writing'', ``human-AI collaboration'', ``generation'', ``evaluation'', and ``LLM'' across the ACL Anthology and ACM Digital Library. This strategy yielded 168 relevant papers, with an initial focus that revealed a notable lack of metrics to assess the intricacies of human-AI writing collaboration. Consequently, we filtered out ACL papers which focus on NLG methods but narrowed our analysis to the 62 ACM papers to further investigate user-facing intelligent writing systems. Our inclusion criteria was that the paper should have discussed the interaction of the writer and the intelligent assistant in writing task. This led to the exclusion of 36 unrelated papers and the addition of eight system papers identified through a backward search, culminating in an in-depth examination of 34 ACM papers for our subsequent analysis. 

\subsection{Metrics for Writing Process}
\label{sec:metrics}
Transitioning from traditional solitary writing practices, the integration of digital tools into the writing process necessitates a new model to assess the writer's engagement and contribution. A writing task is compromised of how it is written and what information is written. To capture these two types of contributions in writing we adopt concepts from information theory -- \textcolor{entropy_blue}{\textbf{entropy}} and  \textcolor{info_gain_orange}{\textbf{information gain}} -- as metrics to assess human's and agent's respective contributions in a co-writing context (Fig. \ref{fig:updated cognitive process}). This methodology fills the evaluation gap we identified in the previous section and provides a nuanced framework for understanding the intricate collaboration between humans and AI in writing. We describe their theoretical underpinnings below. 

\textbf{\textcolor{entropy_blue}{Entropy}}, particularly in information theory, measures the uncertainty or randomness in data. 
Entropy in information theory is defined as follows~\cite{shannon2001mathematical}:
\[ H(X) = -\sum_{i=1}^{n} P(x_i) \log_b P(x_i) \]
where \( H(X) \) is the entropy of the random variable \( X \), \( P(x_i) \) represents the probability of each outcome \( x_i \), and the base of the logarithm \( b \) determines the unit of entropy.
Entropy in writing mirrors the unpredictability of style and content, indicating the level of variability or uniformity. Initially, writers experiment with various narrative directions, character arcs, and styles, increasing entropy through creative exploration. However, during revision, this entropy decreases as coherence and order are applied. Throughout the writing process, entropy fluctuates, embodying the ongoing struggle between order and chaos, guiding the transformation from rough drafts to refined texts.

\textbf{\textcolor{info_gain_orange}{Information gain}} is a measure of reduction in uncertainty about a variable after observing new information~\cite{azhagusundari2013feature}. 
\[ IG(T, a) = H(T) - H(T|a)\] where $H(T|a)$ is the conditional uncertainty of T given the value of attribute $a$. As we learn more about T (denoted as "$a$"), we decrease the uncertainty associated with $T$, which constitutes gaining information.
In the writing context, information gain quantifies the amount of uncertainty reduced from the reader's perspective. As the narrative unfolds or as arguments and evidence are presented, new information is introduced to the reader. If this new information significantly reduces the reader's uncertainty about the characters, plot, or the argument being made, it can be said that there has been a high information gain. In other words, information gain measures the human's and agent's contribution in changing the content of the writing.

\subsection{Construction of a Two-Dimensional Taxonomy}

Based on the two dimensions of entropy and information gain we identify in Section~\ref{sec:metrics}, we propose a novel taxonomy, \textbf{CoCo Matrix}\footnote{CoCo stands for \textbf{Co}gnitive \textbf{Co}ntribution}, to categorize the human-agent co-writing contributions into four quadrants. Further, we applied our taxonomy to analyze the 34 selected papers from our literature review in section~\ref{sec:lit_review}. 
We coded the papers~(Table~\ref{tab:sim_dif}) by the four quadrants of our CoCo Matrix taxonomy, the three phrases of cognitive writing process, the interactions mechanism, UI design and the system's evaluations. We further discuss our findings in Section~\ref{sec:taxonomy}.

%% file: 05-taxonomy.tex
\section{CoCo Matrix: Taxonomy to Assess Cognitive Contributions}
\label{sec:taxonomy}

The CoCo Matrix, a two-dimensional taxonomy based on varying states of entropy and information gain, serves to evaluate contributions from humans and agents in co-writing across various cognitive stages and writing domains. While the matrix distinguishes between systems that support different types of writing tasks, some systems may fall into multiple quadrants based on the different functions they support. 

\begin{figure*}
    \centering
    \includegraphics[width=1\textwidth]{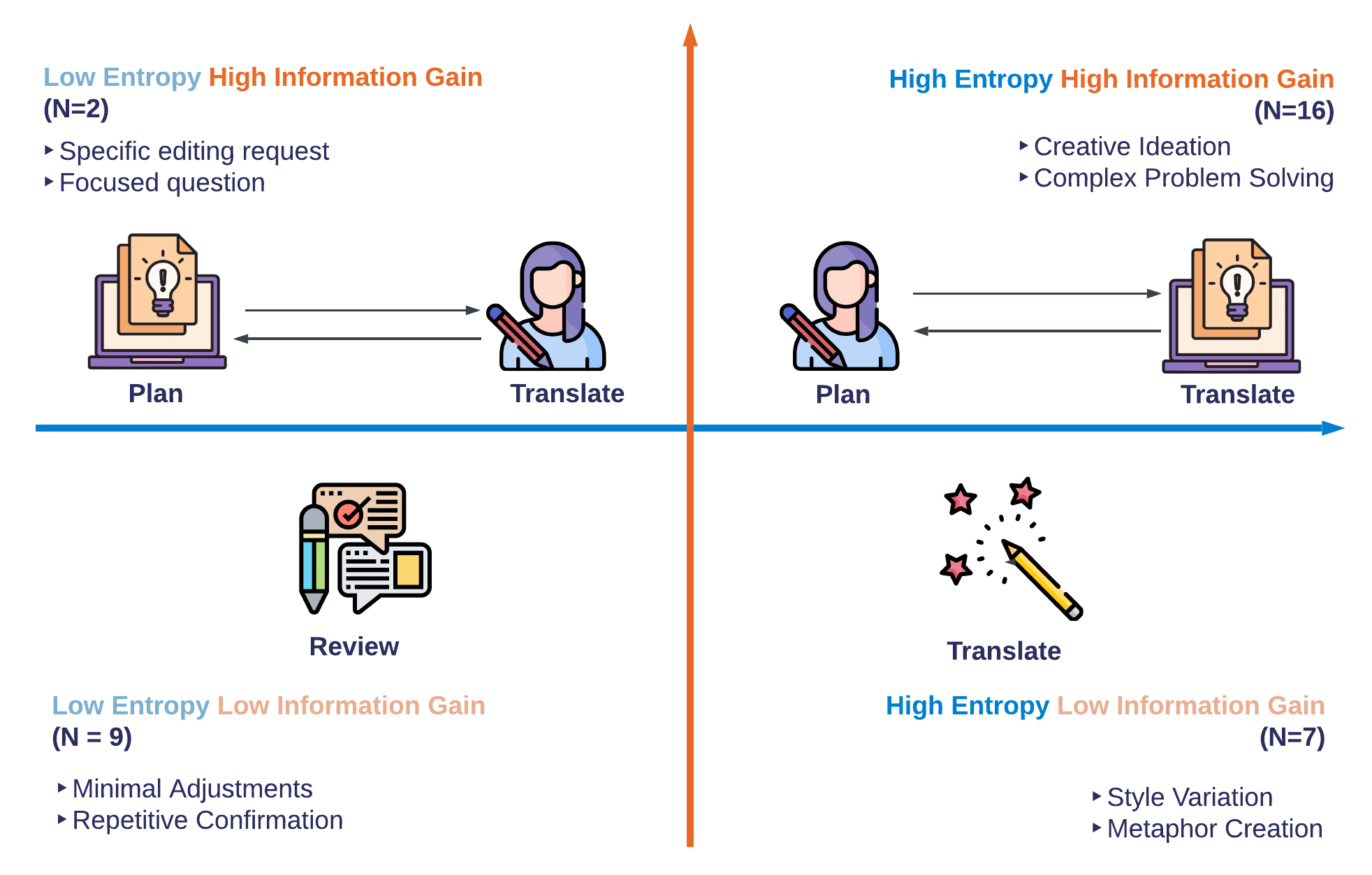}
    \caption{CoCo Matrix - a two-dimensional matrix of entropy and information gain: we analyzed thirty-four systems with our taxonomy. Low entropy and high information gain systems are under-explored.}
    \label{fig:enter-label}
\end{figure*}

\input{Tables/Summary_Table}




\subsection{High entropy, High information gain}
\label{sec:HEHI}

In high entropy, high information gain scenarios, the writing process usually starts with writers providing direction through plans or initial drafts. Agents then provide assistance by expanding these inputs into detailed text, increasing both the diversity and volume of content. This method of collaboration exemplifies a high information gain given the enriched content and a high entropy from the wide range of potential outcomes. 


Writers provide inputs in diverse forms and levels of abstraction, such as drafts ~\cite{zhang2023visar, poddar2023ai_HIHE, dang2023choice}, writing instructions ~\cite{mirowski2023co_HIHE}, or a mix~\cite{chung2022talebrush, yuan2022wordcraft}, expanding the scope of agents for content generation. The agents respond then with direct generation (N = 9) or indirect suggestions (N=7). For example, Talebrush directly generated stories based on a prompt, a character name, and the fortune line given by the users \cite{chung2022talebrush}, while the Choice Over Control study compared four different suggestion interfaces based on the users' diegetic prompts \cite{dang2023choice}. 
\citet{bhat2023interacting_HIHE} showed that agent suggestions would influence the plan of human writers or translation in writing, though they are designed to align with the intentions and goals of writers.
In this quadrant, the focus on the planning (N = 8) and translation (N = 16) stages shows that intelligent agents significantly reduce human writers' effort in translation. This trend reflects the current direction in writing assistants, with nearly half of the reviewed papers (16 out of 34) emphasizing the utility of AI in enhancing writing translation. The reviewed systems in this quadrant tend to influence the intents and decisions of the writer. These systems guide the content creation process by assisting the user in the ideation through planning, and writing through the translation phase. 

\subsection{High entropy, Low information gain}
\label{sec:HELI}

In an interaction dynamic characterized by high entropy and low information gain, writers and intelligent agents collaborate in a landscape focused on open-ended brainstorming and exploratory prompts, such as experimenting with style variations. This approach thrives on generating a wide array of ideas and stylistic approaches, where the primary goal is not the immediate coherence or the distillation of information, but rather the exploration of creative possibilities. 
This writing interaction provides stylistic choices that can be tailored in subsequent phases. This phase is essential for creative breakthroughs, especially in metaphor creation (N = 2) \cite{gero2019metaphoria_LIHE,kim2023metaphorian_LIHE} and style variation (N = 5) \cite{gero2019stylistic_LIHE, ko2022we_LIHE}. It allows writers to explore beyond conventional boundaries and discover unique expressions that resonate more deeply.

Tools like Metaphoria \cite{gero2019metaphoria_LIHE} and Metaphorian \cite{kim2023metaphorian_LIHE} facilitate metaphor generation by producing diverse outcomes based on input keywords, aligning with the author's intent. Similarly, style variation is enhanced by tools such as the Stylistic Thesaurus \cite{gero2019stylistic_LIHE}, We-toon~\cite{ko2022we_LIHE}, and slogan writing tools \cite{clark2018creative}, which tailor content to preferred stylistic choices.
These interactions are particularly valuable in the early stages of writing, aiding in exploring and translating initial ideas. This dynamic is crucial for creative writing, where the need for novelty and divergence is high. It allows writers to push beyond conventional limits to uncover unique expressions and narratives. 

\subsection{Low entropy, High information gain}
\label{sec:LEHI}

In the low entropy, high information gain quadrant, intelligent agents primarily lead the planning phase, setting some creative directions, while writers refine and translate these ideas into polished text based on their own preferences and writing goals. This structure is exemplified by tools like Sparks, which generates sentences to inspire science writing through specific templates \cite{gero2022sparks_HILE}, and Heteroglossia, a crowd-sourcing tool designed as a Google Doc add-on to foster story idea generation from crowd workers via role-play strategies \cite{huang2020heteroglossia_HILE}. 

This quadrant shows the opposite collaboration as the high entropy, high information gain quadrant (Sec \ref{sec:HEHI}): agents initiate with broad idea generation, guiding the writing process's conceptual phase, while writers are responsible for the articulation of these concepts into finalized content. The resulting interaction between users and intelligent agents is defined by focused, deliberate inputs that drive the agents towards producing outputs closely aligned with the writer's specific objectives. This dynamic yields a concentrated, low-entropy creative environment favorable for creating content with high clarity, relevance, and precision. The human writers guide the process, significantly reducing the potential for deviation from the intended narrative or argumentative path, ensuring that each piece of content and every refinement are purposefully directed toward enriching the depth and precision of the writing.

\subsection{Low entropy, Low information gain}
\label{sec:LELI}

This interaction type, marked by low entropy and low information gain, focuses on the meticulous aspects of the editing process, such as formatting checks and grammar corrections. 
While these tasks may involve minimal adjustments and repetitive confirmations, they are fundamental to ensuring the clarity, coherence, and professionalism of the final text. This controlled and predictable environment allows for high accuracy and consistency, which are essential in formal writing, academic publications, and professional documentation. Although this phase may not contribute significantly to novelty or content depth, it is indispensable to polishing and refining the document, ensuring that creative and intellectual efforts are communicated and free from distractions. Recognizing the importance of this phase, strategies can be integrated to streamline these tasks efficiently, potentially leveraging AI capabilities to automate aspects of the process, thereby freeing human creativity for more complex and nuanced tasks.

Research in this category primarily supports the review phases (N = 9). For instance, the use of adaptive nudging to facilitate argumentation self-evaluation by highlighting aspects such as claims and readability offers targeted improvements in argumentative writing \cite{wambsganss2020adaptive_LILE,wambsganss2022improving_LILE}. In addition, tools that provide readability feedback \cite{karolus2023your_LILE} or continuous text summaries \cite{dang2022beyond} aid writers by offering real-time insights into the clarity and coherence of their text.
These interactions, which occur predominantly in the review phase, are particularly aligned with argumentative writing \cite{wambsganss2020adaptive_LILE,wambsganss2022improving_LILE}. This type of writing demands a rigorous presentation and benefits from the precision offered by AI-assisted review processes. 



%% file: Tables/Summary_Table.tex
\begin{table*}[!ht]
    \centering
    \setlength{\tabcolsep}{10pt} 
    \renewcommand{\arraystretch}{2} 
    \begin{tabular}{| l | l  l|} 
    \hline
    \textbf{Taxonomy}     & \textbf{Writing Process} & \textbf{Paper}\\ 
    \hline
    \multirow{7}{*}{\makecell[l]{ High Entropy,\\ High Information Gain\\(N=16)}}
        
        & Plan, Translate, Review 
            & \makecell[l]{
            VISAR \cite{zhang2023visar}\\
            Talebrush \cite{chung2022talebrush} \\ 
            Academic Writing \cite{buruk2023academic}
            }\\
        \cline{2-3}
        & Plan, Translate
                &\makecell[l]{
            Leaps \cite{singh2023hide_HIHE}\\
            Dramatron \cite{mirowski2023co_HIHE}\\
            BunCho \cite{osone2021buncho}\\
            Next-phrase suggestion \cite{bhat2023interacting_HIHE}\\
            Narratron \cite{zhao2023narratron_HIHE}
            }\\
        \cline{2-3}
        & Translate, Review
            &\makecell[l]{
            Writing with Machine (Story) \cite{clark2018creative}    
            }\\        
        \cline{2-3}
        & Translate
            & \makecell[l]{
            Self-presentation \cite{poddar2023ai_HIHE}\\
            Word-craft \cite{yuan2022wordcraft}\\
            Choice Over Control \cite{dang2023choice}\\
            Parallel Phrase Suggestion \cite{buschek2021impact_HIHE}\\
            Writing with RNN \cite{roemmele2018automated_HIHE}\\
            LyriSys \cite{watanabe2017lyrisys_HIHE}\\
            Redhead \cite{ghajargar2022redhead} 
            }\\
        
    \hline
    
    \multirow{4}{*}{\makecell[l]{High Entropy, \\Low Information Gain\\(N=7)}} 
        &  Plan, Translate, Review  
            & \makecell[l]{
            Academic Writing \cite{buruk2023academic}\\
            Metaphorian \cite{kim2023metaphorian_LIHE}
            }\\
        \cline{2-3}
        & Translate, Review 
            & \makecell[l]{
            Writing with Machine (Slogan) \cite{clark2018creative} 
            }\\
        \cline{2-3}
        & Plan
            &\makecell[l]{
            Stylistic Thesaurus \cite{gero2019stylistic_LIHE}\\
            Metaphoria \cite{gero2019metaphoria_LIHE}
                }\\
        \cline{2-3}
        & Translate
            & \makecell[l]{
            We-toon \cite{ko2022we_LIHE}\\
            Redhead \cite{ghajargar2022redhead}
            }\\
    \hline
    \makecell[l]{ Low Entropy, \\High Information Gain\\(N=2)} 
        & Plan, Translate 
            & \makecell[l]{
            Sparks \cite{gero2022sparks_HILE} \\
            Heteroglossia \cite{huang2020heteroglossia_HILE}
            }\\
    
    \hline
    \multirow{6}{*}{\makecell[l]{Low Entropy,\\Low Information Gain\\(N=9)} }
        & Plan, Translate, Review 
            & \makecell[l]{
            Academic Writing \cite{buruk2023academic}
            }\\
        \cline{2-3}
        & Translate, Review
            & \makecell[l]{
            AmbientLetter \cite{toyozaki2018ambientletter_LILE}\\
            Social Media Writing \cite{wu2019design_LILE}
            }\\
        \cline{2-3}
        & Review 
            & \makecell[l]{
            Automatic Writing Feedback \cite{chang2021exploring_LILE}\\
            Argumentation Self-Evaluation Nudging \cite{wambsganss2022improving_LILE}\\
            Readability Awareness \cite{karolus2023your_LILE}\\
            Continuous Text Summaries \cite{dang2022beyond_LILE}\\
            Adaptive Learning for Argumentation \cite{wambsganss2020adaptive_LILE}\\
            Literary Style Tool \cite{sterman2020interacting_LILE}
            }\\
    \hline
    \end{tabular}
    \caption{Summary of thirty-four papers analyzed }

    \label{tab:sim_dif}
\end{table*}

%% file: 06-Discussion.tex
\section{Discussion}
Our framework introduces a methodological approach to analyzing human-agent co-writing, emphasizing the significance of entropy and information gain. This perspective illuminates the intricacies of collaboration between human creativity and AI, which is a pivotal guide for enhancing the co-writing process. By categorizing interactions into distinct dimensions, we have mapped out a landscape where an intelligent agent's role is to augment, not overshadow, human originality and quality.
Our taxonomy contributes to a clearer understanding of the co-writing process and is a reference point for future technology advancements. Our research shows that most systems fall into the high entropy, high information gain category (N = 16), easing the integration of AI agents. We observe these systems have been dominant in last two years, while low entropy and low information gain systems are mostly older systems. This shift reflects the rapid advancements in LLMs  and signals a new era of our interaction with machines. The next common quadrant is low entropy and low information gain (N = 9), noted for their reliability and straightforward development. Meanwhile, high entropy and low information gain (N = 7) interactions are pivotal for expanding creativity, though they are less common. Significantly under-explored, yet highly promising, are low entropy and high information gain interactions (N = 2), suggesting an opportunity for future exploration to leverage AI more effectively and non-intrusively.

Our findings guide developers and researchers in creating intelligent writing tools that enhance the writer's voice and intent. By focusing on under-explored areas of low entropy and high information gain interactions, there is vast potential to innovate in ways that respect human agency and maximize the benefits of AI collaboration. 

Overall, CoCo Matrix comprehensively categorizes four types of interaction by entropy and information gain of continuous minimal changes, which highlights the contribution of the writer's cognition through the writing process.

